\documentclass[useAMS,usenatbib]{mn2e}
\usepackage{graphicx}
\usepackage{amssymb}
\usepackage{color}

\title[The multi-band nonthermal emission from RX J1713.7-3946]{The multi-band
nonthermal emission from the supernova remnant RX J1713.7-3946}
\author[Fang et al.]{J. Fang \thanks{email: fangjun1653@126.com}, L. Zhang, J.F. Zhang, Y.Y. Tang, H. Yu\\
Department of Physics, Yunnan University, Kunming, China}

\begin{document}
\pagerange{\pageref{firstpage}--\pageref{lastpage}} \pubyear{2008}

\maketitle

\label{firstpage}

\begin{abstract}
Nonthermal X-rays and very high-energy (VHE) $\gamma$-rays have been
detected from the supernova remnant (SNR) RX J1713.7-3946, and
especially the recent observations with the \textit{Suzaku}
satellite clearly reveal a spectral cutoff in the X-ray spectrum,
which directly relates to the cutoff of the energy spectrum of the
parent electrons. However, whether the origin of the VHE
$\gamma$-rays from the SNR is hadronic or leptonic is still in
debate. We studied the multi-band nonthermal emission from RX
J1713.7-3946 based on a semi-analytical approach to the nonlinear
shock acceleration process by including the contribution of the
accelerated electrons to the nonthermal radiation. The results show
that the multi-band observations on RX J1713.7-3946 can be well
explained in the model with appropriate parameters and the TeV
$\gamma$-rays have hadronic origin, i.e., they are produced via
proton-proton (p-p) interactions as the relativistic protons
accelerated at the shock collide with the ambient matter.
\end{abstract}

\begin{keywords}
radiation mechanisms: non-thermal -- supernova remnants  --
gamma-rays: theory -- ISM: individual(RX J1713.7-3946)
\end{keywords}

\section{Introduction}

Supernova remnants (SNRs) are broadly believed to be acceleration
sites of the Galactic cosmic rays. X-ray observations can provide an
indication of electrons being accelerated up to multi-TeV energies
and very high-energy (VHE) observations indicate that the particles
would be accelerated up to hundreds of TeV or more in the SNRs.
However, the origin of the VHE $\gamma$-rays from SNRs is still
uncertain because the TeV $\gamma$-rays from each source can usually
be explained either by hadronic models in which they are produced
via p-p interactions or by leptonic ones in which they are from
inverse Compton scattering of the parent electrons
\citep[e.g.,][]{BV06,A07,ZF07,ZF08,Y06,FZ08,F08,T08,L08}. Detailed
multi-band observations in X-rays, TeV $\gamma$-rays and especially
in the MeV/GeV band with \textit{the Fermi Gamma-ray Space Telescope
(FGST)} satellite are important to eliminate the uncertainty on the
origin of TeV $\gamma$-rays from SNRs.

RX J1713.7-3946 (G347.3-0.5) is a shell-type SNR with faint radio
emission \citep{L04} and strong nonthermal X-ray emission
\citep{S99}. TeV $\gamma$-rays from the SNR shell were first
detected by \textit{CANGAROO} \citep{M00} and confirmed by
\textit{CANGAROO-II} \citep{E02} and \textit{H.E.S.S.} \citep{A06}.
Recently, \citet{A07} reported the three-year \textit{H.E.S.S.}
observations on the remnant. The combined data significantly
increase statistics and the VHE spectrum extends over three orders
of magnitude up to energies $\sim100$ TeV. The X-ray observations on
the SNR have been performed with \textit{Suzaku} and a wide-band
X-ray spectrum (0.4--40 keV) with high statistics has been shown by
combining the X-ray Imaging Spectrometer (XIS) and the Hard X-ray
Detector (HXD) spectrum \citep{Tak08,T08}. The X-ray spectrum shows
a clear cutoff shape, which allows the energy spectrum of the parent
electrons to be more clearly investigated.

In this paper, we investigate the multi-band nonthermal emission
from RX J1713.7-3946 in the frame of nonlinear diffusive shock
acceleration mechanism, which has been studied extensively both
numerically \citep[e.g.,][]{EBJ96, BV97, KJ06, EPSBG07} and
semi-analytically \citep[e.g.,][]{MD01, BAC07, ABG08}. The origin of
the TeV $\gamma$-rays from the source had been numerically studied
by \citet{BV08} using the nonlinear kinetic theory of cosmic ray
acceleration in SNRs. They shown that the recent high-energy
observations with \textit{H.E.S.S.} and \textit{Suzaku} agree well
with their previous study \citep{BV06} on the SNR and the hadronic
origin of the TeV photons is more favored than the leptonic one.
Moreover, they argued that the non-thermal X-ray emission can
correlate with the $\gamma$-ray one due to the correlation between
the magnetic field amplification with the accelerated nuclear
particles and the associated streaming instabilities \citep[see
details in][]{BV08}. Here we calculate the spectrum of the particles
accelerated at shocks using the semi-analytical method proposed in
\citet{B02} and \citet{BGV05} and assume that the accelerated
electrons have the same spectrum of the protons up to a maximum
energy determined by synchrotron losses \citep{E00}, and then
calculate the multi-band nonthermal photon spectrum from RX
J1713.7-3946. The results show that the multi-band spectrum of the
SNR can be well reproduced with appropriate parameters in the model
and the observed TeV $\gamma$-rays are produced mainly via hadronic
interactions.

Very recently, \citet{MAB08} calculated the high-energy spectrum for
RX J1713.7-3946 in the context of the nonlinear particle
acceleration process at shocks, which is similar to that using in
this paper. However, our approach is different from \citet{MAB08}.
Firstly, the shape of the multi-band spectrum is sensitive to the
Mach number of the shock; therefore, we investigate the nonthermal
spectrum from the point of the Mach number and find the multi-band
observations on the remnant can be well reproduced with appropriate
parameters in the model. Secondly, they treated the cut off in the
spectrum of electrons as $\exp[-(E/E_{e, {\rm max}})^2]$; however,
we find the \textit{Suzaku} observation can still be well reproduced
with the conventional cut off as $\exp[-(E/E_{e, {\rm max}})]$ in
our approach; Finally, the multi-band observations for the SNR
consist well with the model results with the parameters (see
Fig.\ref{fig:spe3n}), $T_0=10^7$ K, which is argued to be the
temperature of the bubble around the SNR, and $M_0=8.0$,
corresponding to a shock speed of $\sim3000$ km s$^{-1}$, whereas a
value of $10^6$ K for $T_0$ is used in \citet{MAB08}.

The structure of this paper is as follows. In \S 2, we briefly
review the model used here and show our calculation results, and
give some discussion and conclusions in \S 3.

\section{The model and results}

The pitch-angle averaged steady-state distribution of the protons
accelerated at a shock in one dimension satisfies the diffusive
transport equation \citep{MD01, B02, ABG08},
\begin{eqnarray}
\nonumber \frac{\partial}{\partial x}\left
[D\frac{\partial}{\partial x}f(x, p)\right ] &-& u\frac{\partial
f(x, p)}{\partial x} \\ &+& \frac{1}{3}\frac{du}{dx}p\frac{\partial
f(x, p)}{\partial p} + Q(x, p) = 0, \label{eq:diff}
\end{eqnarray}
where the coordinate $x$ is directed along the shock normal from
downstream to upstream, $D$ is the diffusion coefficient and $u$ is
the fluid velocity in the shock frame, which equals $u_2$ downstream
($x<0$) and changes continuously upstream, from $u_1$ immediately
upstream ($x=0^+$) of the subshock to $u_0$ at upstream infinity
($x=+\infty$). With the assumption that the particles are injected
at immediate upstream of the subshock, the source function can be
written as $Q(x, p)=Q_0(p)\delta(x)$. For monoenergetic injection,
$Q_0(p)$ is
\begin{equation}
Q_0(p) = \frac{\eta n_{\rm{gas, }1}u_1}{4\pi
p_{\rm{inj}}^2}\delta(p-p_{\rm{inj}})\;\; , \label{eq:Q0}
\end{equation}
where $p_{\rm{inj}}$ is the injection momentum, $n_{\rm{gas, }1}$ is
the gas density at $x=0^+$ and $\eta$ is the fraction of particles
injected in the acceleration process. With the injection recipe
known as thermal leakage, $\eta$ can be described as $\eta =
4(R_{\rm{sub}}-1)\xi^3e^{-\xi^2}/3\pi^{1/2}$ \citep{BGV05, ABG08},
where $R_{\rm{sub}}=u_1/u_2$ is the compression factor at the
subshock and $\xi$ is a parameter of the order of 2--4 describing
the injection momentum of the thermal particles in the downstream
region ($p_{\rm{inj}}=\xi p_{\rm{th,}2}$). We use $\xi=3.5$ as in
\citet{ABG08}, $p_{\rm{th,}2}=(2m_pk_{\rm{B}}T_2)^{1/2}$ is the
thermal peak momentum of the particles in the downstream fluid with
temperature $T_2$, $m_p$ is the proton mass and $k_{\rm{B}}$ is the
Boltzmann constant. Assuming the heating of the gas upstream is
adiabatic, with the conservation condition of momentum fluxes
between the two sides of the subshock, we can derive the relation
between the temperature of the gas far upstream $T_0$ and $T_2$,
i.e., $T_2 = (\gamma_g M_0^2/R_{\rm tot})[R_{\rm sub}/R_{\rm tot} -
1/R_{\rm tot} + (1/\gamma_g M_0^2)(R_{\rm tot}/R_{\rm
sub})^{\gamma_g}]T_0$, where $M_0$ is the fluid Mach number far
upstream, $R_{\rm tot}=u_0/u_2$ is the total compression factor,
$\gamma_g$ is the ratio of specific heats ($\gamma_g=5/3$ for an
ideal gas).

\begin{figure*}
\resizebox{0.8\hsize}{!}{\includegraphics{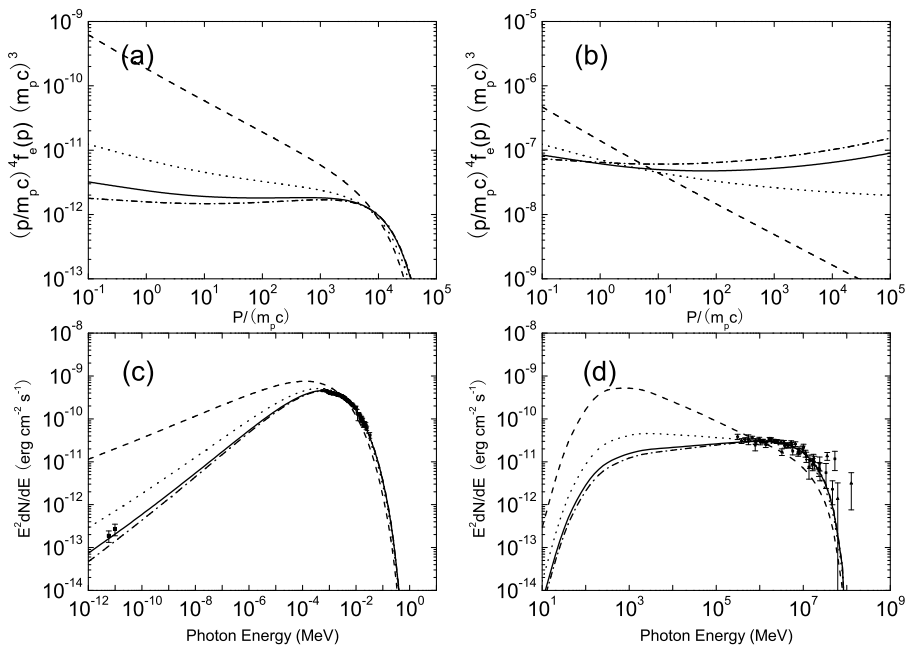}} \caption{The
resulting electron (panel a) and proton (panel b) spectra and the
spectral energy distribution of synchrotron emission of the
electrons (panel c) and p-p interactions (panel d) of the protons
for RX J1713.7-3946 with $d=1$ kpc, $B_{\rm SNR}=100$ $\mu$G,
$p_{\rm max}=1.0\times10^5$ $m_p$c, $E_{{\rm max},e} =10$ TeV for
$M_0=3$, $n_{\rm{gas,}0}=6.05$ cm$^{-3}$, $K_{ep}=1.32\times10^{-3}$
(dashed line); $M_0=5$, $n_{\rm{gas,}0}=0.41$ cm$^{-3}$, $K_{ep}=
1.00\times10^{-4}$ (dotted line); $M_0=8$, $n_{\rm{gas,}0}=0.13$
cm$^{-3}$, $K_{ep}= 3.82\times10^{-5}$ (solid line); and $M_0=12$,
$n_{\rm{gas,}0}=0.08$ cm$^{-3}$, $K_{ep}= 2.44\times10^{-5}$
(dash-dotted line). The ATCA radio data \citep{L04}, \textit{Suzaku}
X-ray data \citep{T08}, and \textit{H.E.S.S.} \citep{A07} data from
the years 2003, 2004 and 2005 are also shown.} \label{fig:spe}
\end{figure*}

RX J1713.7-3946 has been thought as a remnant of type II/Ib
supernova and is evolving in a bubble with a typical temperature
$\geq10^7$ K \citep[e.g.,][]{MAB08}. In this paper, we use
$T_0=10^7$ K, which corresponds to a sound speed of $\sim370$ km
s$^{-1}$, to calculate the multi-band flux for the SNR.

With the assumption that the diffusion is $p$ dependent and
therefore particles with larger momenta move farther away from the
shock than those with lower momenta, only particles with momentum
$\geq p$ can reach the point $x_p$ and thus the pressure of the
accelerated particles at the point can be described as \citep{B02}
\begin{equation}
P_{{\rm{CR}}, p}=\frac{4\pi}{3}\int^{p_{\rm max}}_{p}dp' p'^3 v(p')
f_0(p') \;\; , \label{PCR}
\end{equation}
where $v(p)$ is the velocity of particles with momentum $p$.
Furthermore, the particle distribution function $f_0(p)$ at the
shock can be implicitly written as \citep{B02}
\begin{eqnarray}
\nonumber f_0(p) & = & \left[\frac{3R_{\rm{tot}}}{R_{\rm{tot}}U(p) -
1} \right]\frac{\eta n_{\rm{gas, }0}}{4\pi p_{\rm{inj}}^3}
\\ && \times \exp\left[ -
\int^{p_{\rm{max}}}_{p_{\rm{inj}}}\frac{dp'}{p'}\frac{3R_{\rm
tot}U(p')}{R_{\rm{tot}}U(p') - 1}\right], \label{eq:f0}
\end{eqnarray}
where $n_{\rm{gas, }0}$ is the gas density far upstream
($x=+\infty$), $p_{\rm max}$ is the maximum momentum of the
accelerated particles. We use $p_{\rm max}\sim1.0\times10^5$ $m_p$c
in this paper, which is a typical value for an SNR with ambient
magnetic strength $\sim100$ $\mu$G with an average shock speed of
$\sim5000$ km s$^{-1}$ for a young SNR with an age of $\sim1000$ yr
\citep{Y08}. $U(p)$ can be solved using an equation deduced from the
conservation of the mass and momentum fluxes with the boundary
condition $U(p_{\rm inj}) = R_{\rm sub}/R_{\rm tot}$ and $U(p_{\rm
max})=1$, and then a value of $R_{\rm sub}$ can be achieved by an
iterative procedure to satisfy the boundary conditions \citep{B02}.

Electrons accelerated at a shock are usually treated as test
particles because they carry little momentum and have little
influence on the shock structure. Therefore, their distribution can
not be obtained by general considering momentum and energy
conservations. However, electrons and protons have the same
acceleration rate if they have the same upstream diffusion length,
so the electron and proton spectral shapes should be similar at
superthermal energies \citep{E00}. As a result, the electrons have
the same spectrum of the protons up to a maximum energy determined
by synchrotron losses, and in this paper we simply use
\begin{equation}
f_e(x, p)=K_{ep}f(x, p)\exp(-E(p)/E_{{\rm max}, e}), \label{fe}
\end{equation}
where $E(p)$ is the kinetic energy of the electrons, $E_{{\rm max},
e}$ is the cutoff energy due to the synchrotron losses, and the
electron/proton ratio $K_{ep}$ is treated as a parameter.

Assuming the accelerated particles distribute homogeneously and most
of the emission is from downstream of the shock, and using the
distribution function at the shock to represent the particle
distribution in the whole emitting zone, the volume-averaged
emissivity for photons produced via p-p interactions can be written
as
\begin{equation}
Q(E)=4\pi n_{\rm gas}\int dE_{\rm p} J_{\rm p}(E_{\rm
p})\frac{d\sigma(E, E_{\rm p})}{dE} \;\; , \label{eq:PP}
\end{equation}
where $E_{\rm p}$ is the proton kinetic energy, $n_{\rm gas}$ is the
ambient gas number density, and $J_{\rm p}(E_{\rm p})=v p^2 f_0(p)
dp/dE_{\rm p}$ is the volume-average proton density and $v$ is the
particles' velocity. We use the differential cross-section for
photons $d\sigma(E, E_{\rm p})/dE$ presented in \citet{Ka06} to
calculate the hadronic $\gamma$-rays produced via p-p collisions.
Finally, the photon flux observed at the earth can be obtained with
\begin{equation}
F(E)=\frac{VQ(E)}{4\pi d^2} \;\; , \label{eq:Flux}
\end{equation}
where $d$ is the distance from the earth to the source and $V$ is
the average emitting volume of the source and can be estimated by
$V\approx(4\pi/3)R_{\rm SNR}^3 /R_{\rm tot}$, here $R_{\rm SNR}$ is
the radius of the SNR \citep{E00}. Note that the emission from
secondary $e^{\pm}$ pairs produced from p-p collisions is usually
orders of magnitude smaller than that from the primary electrons and
protons for a young SNR \citep{FZ08}, thus we neglect the component
from the secondary $e^{\pm}$ pairs.

The distance of RX J1713.7-3946 is uncertain and has been revised
sometimes \citep{A06}. The \textit{ASCA} X-ray observation
\citep{K97} and NANTEN CO data \citep{F03,M05} indicate the SNR has
a distance of $\sim 1$ kpc corresponding to an age of $\sim1000$ yr
and we use this value in this paper. The magnetic field in the
diffuse regions where the bulk of the synchrotron emission is
produced is $\geq100$ $\mu$G in order to explain the X-ray flux
ratio between the diffuse and compact regions \citep{U07,T08}. We
use $100$ $\mu$G to calculate the synchrotron emission of the
accelerated electrons and the cutoff energy of the electrons must be
$\sim10$ TeV to well reproduce the X-ray spectrum observed with
\textit{Suzaku}.

Fig.\ref{fig:spe} illustrates the resulting electron, proton and
photon spectra for the SNR RX J1713.7-3946 with $d=1$ kpc, $B_{\rm
SNR}=100$ $\mu$G, $p_{\rm max}=1.0\times10^5$ $m_p$c, $E_{{\rm
max},e} =10$ TeV for $M_0=3$, $n_{\rm{gas,}0}=6.05$ cm$^{-3}$,
$K_{ep}=1.32\times10^{-3}$ (dashed line); $M_0=5$,
$n_{\rm{gas,}0}=0.41$ cm$^{-3}$, $K_{ep}= 1.00\times10^{-4}$ (dotted
line); $M_0=8$, $n_{\rm{gas,}0}=0.13$ cm$^{-3}$, $K_{ep}=
3.82\times10^{-5}$ (solid line); and $M_0=12$, $n_{\rm{gas,}0}=0.08$
cm$^{-3}$, $K_{ep}= 2.44\times10^{-5}$ (dash-dotted line). For each
Mach number, $n_{\rm{gas,}0}$ is normalized to make the resulting
flux at 1 TeV consistent with the \textit{H.E.S.S.} observation, and
$K_{ep}$ is set according to the \textit{Suzaku} results at 2.6 keV.
The ATCA radio data \citep{L04}, \textit{Suzaku} X-ray data
\citep{T08}, and \textit{H.E.S.S.} \citep{A07} data from the years
2003, 2004 and 2005 are also shown. Note that the radio data given
by \citet{L04} is just for the northwest part of the shell and the
data points are multiplied by a factor of two to account for the
emission for the whole remnant as in \citet{A06}. The concave shape
of the particle distribution is more obvious for a higher Mach
number and the spectrum is harder at high momenta; importantly, the
multi-band observed spectrum can be well reproduced with $M_0=8$.

\begin{figure}
\resizebox{\hsize}{!}{\includegraphics{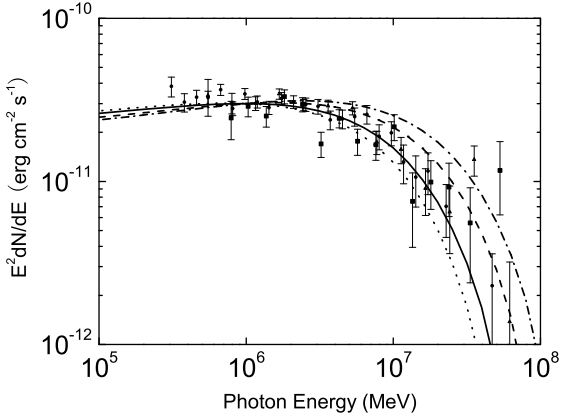}}
\caption{Comparisons of the resulting spectral energy distributions
of p-p interactions for RX J1713.7-3946 for $M_0=8$, $p_{\rm max}=
0.8\times10^5$ $m_{\rm p}c$ (dotted line); $p_{\rm max}=
1.0\times10^5$ $m_{\rm p}c$ (solid line); $p_{\rm max}=
1.5\times10^5$ $m_{\rm p}c$ (dashed line); $p_{\rm max}=
2.0\times10^5$ $m_{\rm p}c$ (dash-dotted line) with the
\textit{H.E.S.S.} data points \citep{A07}.} \label{fig:spe2n}
\end{figure}

The maximum energy of the accelerated electrons for the SNR RX
J1713.7-3946 is limited due to the strong synchrotron losses in our
model, so the shape of the synchrotron emission is nearly constant
with different maximum energies of the protons. The influence of
$p_{\rm max}$ on the spectrum of p-p collisions is shown in
Fig.\ref{fig:spe2n}. Obviously, the TeV observations can be well
reproduced with $p_{\rm max}$ around $1.0\times10^5$ $m_p$c and the
spectrum with $p_{\rm max}= 2.0\times10^5$ $m_{\rm p}c$ relatively
deviates with the bulk of the \textit{H.E.S.S.} data points.
Therefore the maximum energy of the protons accelerated in the SNR
is about 100 TeV. In the following, we use $p_{\rm max}=
1.3\times10^5$ $m_{\rm p}c$ to calculate the multi-band spectrum for
the SNR RX J1713.7-3946, and a similar value of $1.26\times10^5$
$m_{\rm p}c$ was chosen to best fit the high-energy observation with
\textit{H.E.S.S.} by \citet{MAB08}.

\begin{figure}
\resizebox{\hsize}{!}{\includegraphics{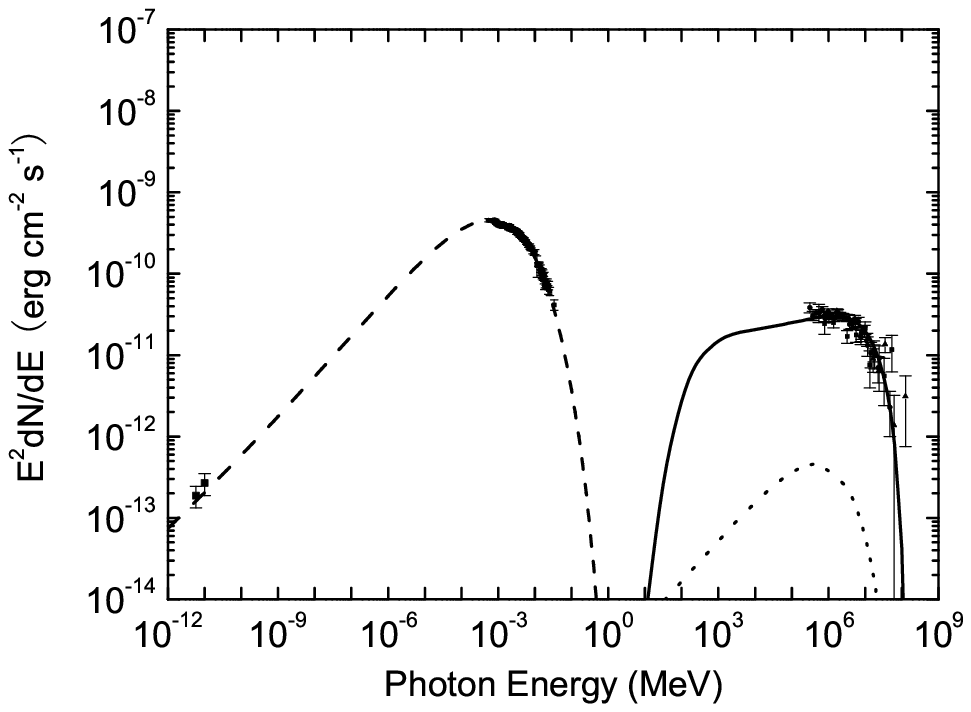}} \caption{The
resulting spectral energy distribution of synchrotron emission
(dashed line), inverse Compton scattering (dotted line) and p-p
interactions (solid line) for RX J1713.7-3946 with $M_0=8$, $p_{\rm
max}= 1.3\times10^5$ $m_{\rm p}c$, $n_{\rm{gas,}0}=0.12$ cm$^{-3}$,
$K_{ep}= 3.92\times10^{-5}$. Others are the same as
Fig.\ref{fig:spe}.} \label{fig:spe3n}
\end{figure}

\begin{figure}
\resizebox{\hsize}{!}{\includegraphics{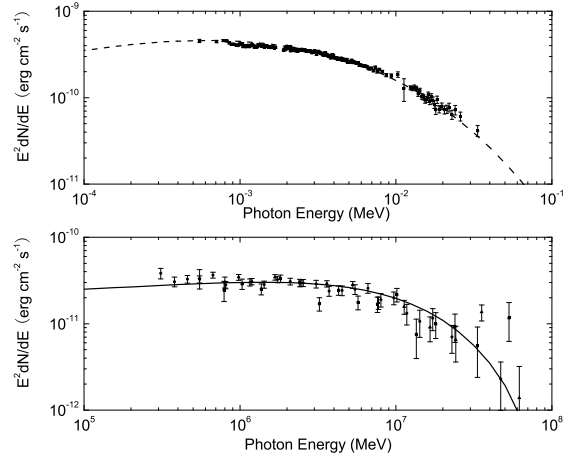}} \caption{Clear
comparisons of the model results with the observed data with
\textit{Suzaku} (upper panel) and \textit{H.E.S.S.}. Others are the
same as Fig.\ref{fig:spe3n}.} \label{fig:spe4n}
\end{figure}

Fig.\ref{fig:spe3n} shows the comparison of the calculated results
for $M_0=8$ with the multi-band observations and Fig.\ref{fig:spe4n}
gives a clear version of the comparisons of the model results with
the \textit{Suzaku} and \textit{H.E.S.S.} observations. The
multi-band nonthermal observations by ATCA, \textit{Suzaku} and
\textit{H.E.S.S.} can be well explained by the model with these
parameters. In calculating the inverse Compston emission of the
accelerated electrons, we use the canonical interstellar values for
the seed photon densities as in \citet{A06}, i.e., 0.25 eV cm$^{-3}$
for the cosmic microwave background, 0.5 eV cm$^{-3}$ for optical
star light and 0.5 eV cm$^{-3}$ for infrared background light. The
contribution of the inverse Compton scattering to the final flux in
the TeV band is orders of magnitude smaller than that of p-p
interactions, and the contribution of Bremsstrahlung is negligible
even compared with the inverse Compton scattering. Obviously, the
observed TeV $\gamma$-rays are from hadronic interactions as the
relativistic protons collide with the ambient matter.

\section{Discussion  and conclusions}

RX J1713.7-3946 is a well-observed shell-type SNR in the radio,
X-ray and TeV $\gamma$-ray bands. Relatively strong magnetic field
with strength $\sim100$ $\mu$G for the regions where the bulk of
synchrotron emission is produced has been deduced from the
observation in X-rays. Especially, the observation with
\textit{Suzaku} gives a wide-band spectrum with energies from 0.4 to
40 keV, and the clear cutoff shape of the observed X-ray spectrum
can constrain the model parameters better than before. Based on the
semi-analytical approach to the particle acceleration process at a
shock, we investigated the multi-band nonthermal spectrum for the
SNR by taking the contribution of electrons into account. The
observed multi-band nonthermal spectrum for the SNR can be well
reproduced in the model with $K_{ep}= 3.92\times10^{-4}$, $M_0=8.0$,
corresponding to a shock speed of $\sim3000$ km s$^{-1}$ for
$T_0=10^7$ K, $p_{\rm max}=1.3\times10^5$ $m_p$c, and the maximum
energy of electrons due to the synchrotron loss is $\sim 10$ TeV in
order to make the model result consistent with the X-ray
observation. The corresponding energy contained by the accelerated
protons is $\sim3\times10^{50}$ erg, and about $15\%$ of the
explosion energy is transferred to protons with a usual explosion
energy of $\sim2\times10^{51}$ erg \citep[e.g.,][]{BV06}. The
results show that the TeV $\gamma$-rays observed by
\textit{H.E.S.S.} are produced predominately via p-p interactions
and the observed X-ray spectrum with \textit{Suzaku} can be well
explained as the synchrotron emission from the accelerated electrons
(see Fig.\ref{fig:spe4n}).

With $n_{\rm{gas,}0}=0.12$ cm$^{-3}$ and $T_0=10^7$ K for a young
SNR with an age of $\sim1000$ yr, thermal X-rays from hot electrons
should be detected from the remnant assuming that electrons and
protons are in thermal equilibrium \citep[e.g.,][]{KW08} and the
lack of thermal X-ray emission from RX J1713.7-3946 seems to
conflict with the above scenario in which the observed TeV
$\gamma$-rays predominately have hadronic origin. However, the
conflict can be excluded by assuming the electrons and protons are
not in equilibrium and the electron temperature is significantly
smaller than the proton's \citep{MAB08}.

\begin{figure}
\resizebox{\hsize}{!}{\includegraphics{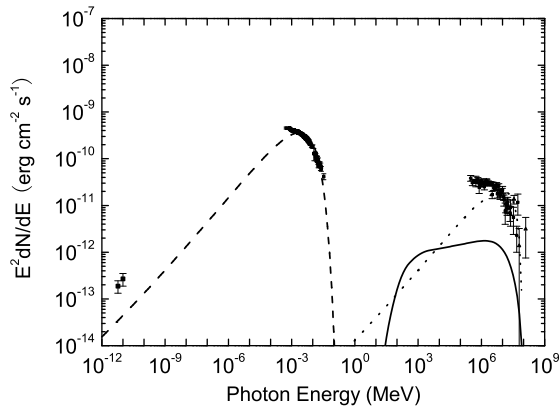}} \caption{The
resulting spectral energy distribution of synchrotron emission
(dashed line), inverse Compton scattering (dotted line) and p-p
interactions (solid line) for RX J1713.7-3946 for $B_{\rm SNR}=10$
$\mu$G, $E_{{\rm max},e} =100$ TeV, $n_{\rm{gas,}0}=0.03$ cm$^{-3}$,
$K_{ep}= 1.23\times10^{-3}$. Others are the same as
Fig.\ref{fig:spe3n}.} \label{fig:spe5n}
\end{figure}

The possibility of leptonic origin of the TeV $\gamma$-rays has been
investigated by many others. For example, \citet{A06} and
\citet{Y08} found that the flat TeV emission detected by
\textit{H.E.S.S.} can not be reproduced with the one-zone model in
which the electron acceleration and gamma-ray emission take place in
the same region unless a relatively large energy density of the
ambient soft field is used in the calculation. Alternatively,
another population of electrons is necessary to reproduce the
multi-band observations with inverse Compton scattering as the
origin of the TeV $\gamma$-rays with relatively low magnetic field
strength $\sim10$ $\mu$G \citep{T08, Y08}, which is much less than
the value deduced by \citet{U07} and \citet{T08}.

Now, we investigate the possible leptonic origin of the TeV photons
with the model in Sec. 2. Fig.\ref{fig:spe5n} shows the resulting
spectrum for $B_{\rm SNR}=10$ $\mu$G, $E_{{\rm max},e} =100$ TeV,
$n_{\rm{gas,}0}=0.03$ cm$^{-3}$, and $K_{ep}= 1.23\times10^{-3}$
which is normalized according to the \textit{Suzaku} observation at
2.6 keV. With such a weak magnetic field, as obtained before, the
inverse Compton emission can not explain the flat TeV spectrum
observed with \textit{H.E.S.S.}; moreover, the synchrotron emission
of electrons can not well reproduce the \textit{Suzaku} spectrum in
the X-ray band and deviates the ATCA radio data.

In summary, the multi-wavelength spectrum of the SNR RX J1713.7-3946
from radio to $\gamma$-ray bands can be well modeled with $B_{\rm
SNR}\sim100$ $\mu$G in the semi-analytical nonlinear case. VHE
$\gamma$-rays from the SNR are produced predominately via the
$\pi^0$ decay in p-p collisions and the X-ray spectrum obtained with
\textit{Suzaku} can be well explained as the synchrotron emission of
the accelerated electrons. On the other hand, the radio, X-ray and
VHE $\gamma$-ray observations can not be well explained in the case
of leptonic origin of TeV $\gamma$-rays with lower magnetic field
strength $\sim10$ $\mu$G using the model in this paper.

\section*{Acknowledgements}
This work is partially supported by a Distinguished Young Scientists
grant from the National Natural Science Foundation of China (NSFC
10425314), NSFC grant 10778702, NSFC grant 10778726, and a 973
Program (2009CB824800).

\label{lastpage}

\end{document}